\documentclass[conference]{IEEEtran}
\usepackage{pstricks,pst-node,pst-text,pst-3d}
\usepackage[utf8]{inputenc}
\usepackage{psfrag}
\usepackage{graphicx}

\def\3lrBL{$SU(3)_{c}\otimes SU(2)_L \otimes SU(2)_R \otimes U(1)_{B-L}$}
\def\10{SO(10)}
\def\vev#1{\left\langle #1\right\rangle}
\def\vb#1{\vbox to #1 pt{}}
\def\21{SU(2) $\otimes$ U(1) }
\def\bold#1{\setbox0=\hbox{$#1$} 
     \kern-.025em\copy0\kern-\wd0 
     \kern.05em\copy0\kern-\wd0 
     \kern-.025em\raise.0433em\box0 }
\DeclareMathAlphabet{\mathsc}{OT1}{cmr}{m}{sc}

\begin{document}

\title{Supersymmetric Models for Neutrino Mass}

\author{\authorblockN{Jorge C. Rom\~ao}
\authorblockA{Instituto Superior T\'ecnico\\
Departamento de F\'{\i}sica and CFTP\\
A. Rovisco Pais 1, 1049-001 Lisboa, Portugal\\
Email: jorge.romao@ist.utl.pt}
}

\maketitle

\begin{abstract}
We review models for neutrino mass, with special emphasis in 
supersymmetric models where R--parity is broken either
explicitly or spontaneously.  The simplest unified extension of the
\texttt{MSSM} with explicit bilinear R--parity violation provides a predictive
scheme for neutrino masses and mixings which can account for the
observed atmospheric and solar neutrino anomalies.  Despite the
smallness of neutrino masses R-parity violation is observable at
present and future high-energy colliders, providing an unambiguous
cross-check of the model. This model can be shown to be an effective
model for the, more theoretically satisfying, spontaneous broken
theory. The main difference in this last case is the appearance of a
massless particle, the majoron, that can modify the decay modes of the
Higgs boson, making it decay invisibly most of the time.
\end{abstract}

% no keywords

\IEEEpeerreviewmaketitle

\section{Introduction}

Despite the tremendous effort that has led to the discovery of
neutrino mass~\cite{fukuda:1998mi,ahmad:2002jz,eguchi:2002dm} the
mechanism of neutrino mass generation will remain open for years to
come (a detailed analysis of the three--neutrino oscillation
parameters can be found in ~\cite{maltoni:2004ei}).  The most popular
mechanism to generate neutrino masses is the seesaw
mechanism~\cite{minkowski:1977sc,gell-mann:1980vs,yanagida:1979,mohapatra:1981yp,chikashige:1981ui,schechter:1980gr,lazarides:1980nt,malinsky:2005bi}. 
Although the seesaw fits naturally in SO(10) unification models, we
currently have no clear hints that uniquely point towards any
unification scheme.

Therefore it may well be that neutrino masses arise from physics
having nothing to do with unification, such as certain seesaw
variants~\cite{mohapatra:1986bd}, and models with radiative
generation~\cite{zee:1980ai,babu:1988ki}.
Here we focus on the specific case of low-energy supersymmetry with
violation of R--parity, as the origin of neutrino
mass. R--parity is defined as $R_p = (-1)^{3B+L+2S}$ with $S$,
$B$, $L$ denoting spin, baryon and lepton numbers,
respectively~\cite{aulakh:1982yn}.

In these models R--parity can be broken either explicitly or
spontaneously. In the first case we consider the bilinear R--parity
violation (\textbf{BRpV}) model, the simplest effective description of
R-parity violation~\cite{diaz:1998xc}. The model not only accounts for
the observed pattern of neutrino masses and
mixing~\cite{diaz:2003as,romao:1999up,hirsch:2000ef,hirsch:2004he},
but also makes predictions for the decay branching ratios of the
lightest supersymmetric
particle~\cite{hirsch:2002ys,porod:2000hv,restrepo:2001me,hirsch:2003fe}
from the current measurements of neutrino mixing
angles~\cite{maltoni:2004ei}. In the second case R--parity violation
takes place ``a la Higgs'', i.e., spontaneously, due to non-zero
sneutrino vacuum expectation values
(vevs)~\cite{masiero:1990uj,romao:1992vu,shiraishi:1993di}.
In this case one of the neutral CP-odd scalars is identified with the
majoron, $J$. In contrast with the seesaw majoron, ours is characterized by
a small scale (TeV-like) and carries only one unit of lepton number.
In previous
studies~\cite{romao:1992ex,romao:1992zx,decampos:1994rw,decampos:1996bg}
it was noted that the spontaneously R--parity violation (\textbf{SRpV})
model leads to the possibility of invisibly decaying Higgs bosons,
provided there is an \21 singlet superfield $\Phi$ coupling to the
electroweak doublet Higgses, the same that appears in the
\textbf{NMSSM}. We have reanalyzed \cite{hirsch:2004rw,hirsch:2005wd}
this issue taking into account the small masses indicated by current
neutrino oscillation data~\cite{maltoni:2004ei}.  We have shown
explicitly that the invisible Higgs boson decay Eq.~(\ref{eq:HJJ}),
\begin{equation}
  \label{eq:HJJ}
  h \to J J
\end{equation}
can be the most important mode of Higgs boson decay. This
is remarkable, given the smallness of neutrino masses required to fit
current neutrino oscillation data.

\section{Models for Neutrino Mass}

In 1980, Weinberg\cite{weinberg:1980bf} noticed that the
dimension-five operator
\begin{equation}
  \label{eq:1}
  \mathcal{L}_{\rm Dim5}=L \phi \,  L \phi
\end{equation}
could induce neutrino masses:
\begin{center}
  \psfrag{n}{$\nu_{L}$}
  \psfrag{f}{$\vev{\phi}$}
  \includegraphics[height=25mm]{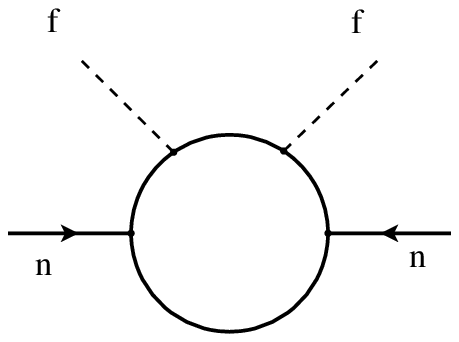}
\end{center}
The models that can lead to this type of operator can be classified in
\textit{Seesaw Models} and \textit{Radiative Models}, that we now
briefly review.

\subsection{Seesaw models for neutrino mass}

\subsubsection{Type I mechanism}

In models with right handed neutrinos
\begin{equation}
  \label{eq:2}
-\mathcal{L}=\overline{\nu_L} m_D \nu_R +  \frac{1}{2}\overline{\nu_L^c} M_R \nu_R  
\end{equation}
where $m_D= Y_{\nu} v$. The seesaw (Type I)~\cite{minkowski:1977sc,gell-mann:1980vs,yanagida:1979,schechter:1980gr,lazarides:1980nt} formula is
\begin{equation}
  \label{eq:3}
m_{\rm eff}^{\rm I}= - (v Y_{\nu}) M_R^{-1} (v Y_{\nu})^T  
\end{equation}
which corresponds to the following diagram:

\begin{center}  
  \psfrag{nL}{$\nu_L$}
  \psfrag{nR}{$\nu_R$}
  \psfrag{MR}{$M_R$}
  \psfrag{f}{$\vev{\phi}$}
\hskip -2mm
 \includegraphics[clip,height=25mm]{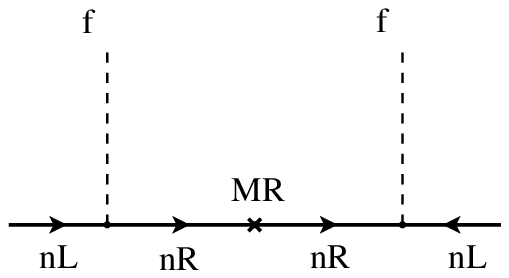}
\end{center}

%%%-------------------- SLIDE -----------------------
\subsubsection{Type II mechanism}

In models with Higgs triplets 
\begin{equation}
-\mathcal{L}=\frac{1}{2} Y_{\Delta} \overline{\nu_L^c} i
  \tau_2 \Delta_L  \nu_L + \mu \phi^T \Delta_L \phi + M^2_{\Delta}
  \Delta_L^{\dagger}   \Delta_L + \cdots 
\end{equation}
we obtain the type II seesaw formula~\cite{mohapatra:1981yp,schechter:1980gr,lazarides:1980nt,schechter:1981cv} 
\begin{equation}
  \label{eq:4}
m_{\rm eff}^{\rm II}=  \frac{v^2 \mu Y_{\Delta}} {M^2_{\Delta}}   
\end{equation}
which corresponds to the following diagram
\begin{center}  
  \psfrag{nL}{$\nu_L$}
  \psfrag{nR}{$\nu_R$}
  \psfrag{MR}{$M_R$}
  \psfrag{f}{$\vev{\phi}$}
  \psfrag{D}{$\Delta^0$}
  \psfrag{m}{$\mu$}
  \psfrag{YD}{$Y_{\Delta}$}
   \includegraphics[height=25mm]{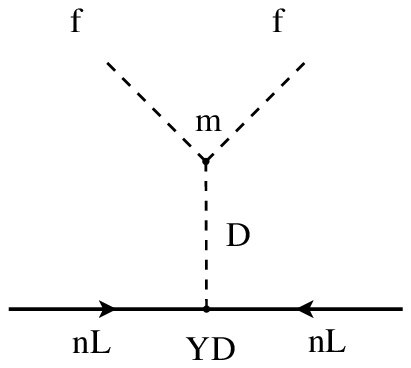}
\end{center}

%%%-------------------- SLIDE -----------------------
\subsubsection{Inverse Seesaw}

In addition to the normal neutrinos $\nu_L$, the inverse
seesaw\cite{mohapatra:1986bd} 
uses two sequential SU(3) $\otimes$ SU(2) $\otimes$ U(1)
singlets $\nu_L^{c}$, $S_L$, corresponding to the following mass matrix
\begin{equation}
  \label{eq:5}
  M_{\nu} =
\left(\begin{array}{ccc}
    0 & Y_{\nu} v& 0 \\
    Y_{\nu}^{T}v & 0 & M\\
    0 & M & \mu
\end{array}\right) 
\end{equation}
where $Y_{\nu}$ are Yukawa couplings, $M$ and $\mu$ are SU(3)
$\otimes$ SU(2) $\otimes$ U(1) invariant mass entries. The effective
mass matrix is then
\begin{equation}
  \label{eq:6}
m_{\rm eff}^{\rm Inv}=  - v^{2}(Y {M^{T}}^{-1}) \mu ({M}^{-1}Y^{T})  
\end{equation}
The smallness of $\mu$ is natural, in t'Hooft's sense. However, there
is no dynamical understanding of this smallness.

%%%-------------------- SLIDE -----------------------
\subsubsection{A Supersymmetric  SO(10) Inverse Seesaw}

\begin{figure}[ht]
  \centering
      \includegraphics[width=0.40\textwidth]{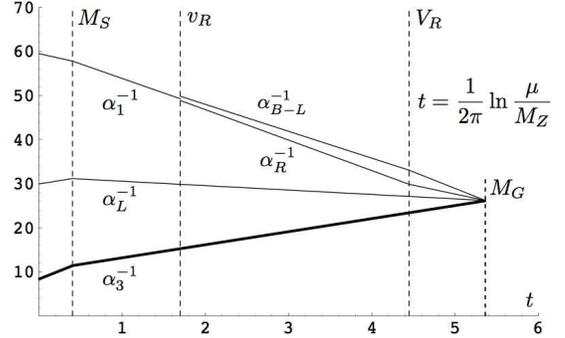}\\      
    \caption{Gauge coupling unification in the in model of
      Ref.\cite{malinsky:2005bi}.} 
    \label{fig:1}
  \end{figure}

\begin{figure}
  \centering
    \begin{tabular}{c}
      \includegraphics[width=0.45\textwidth]{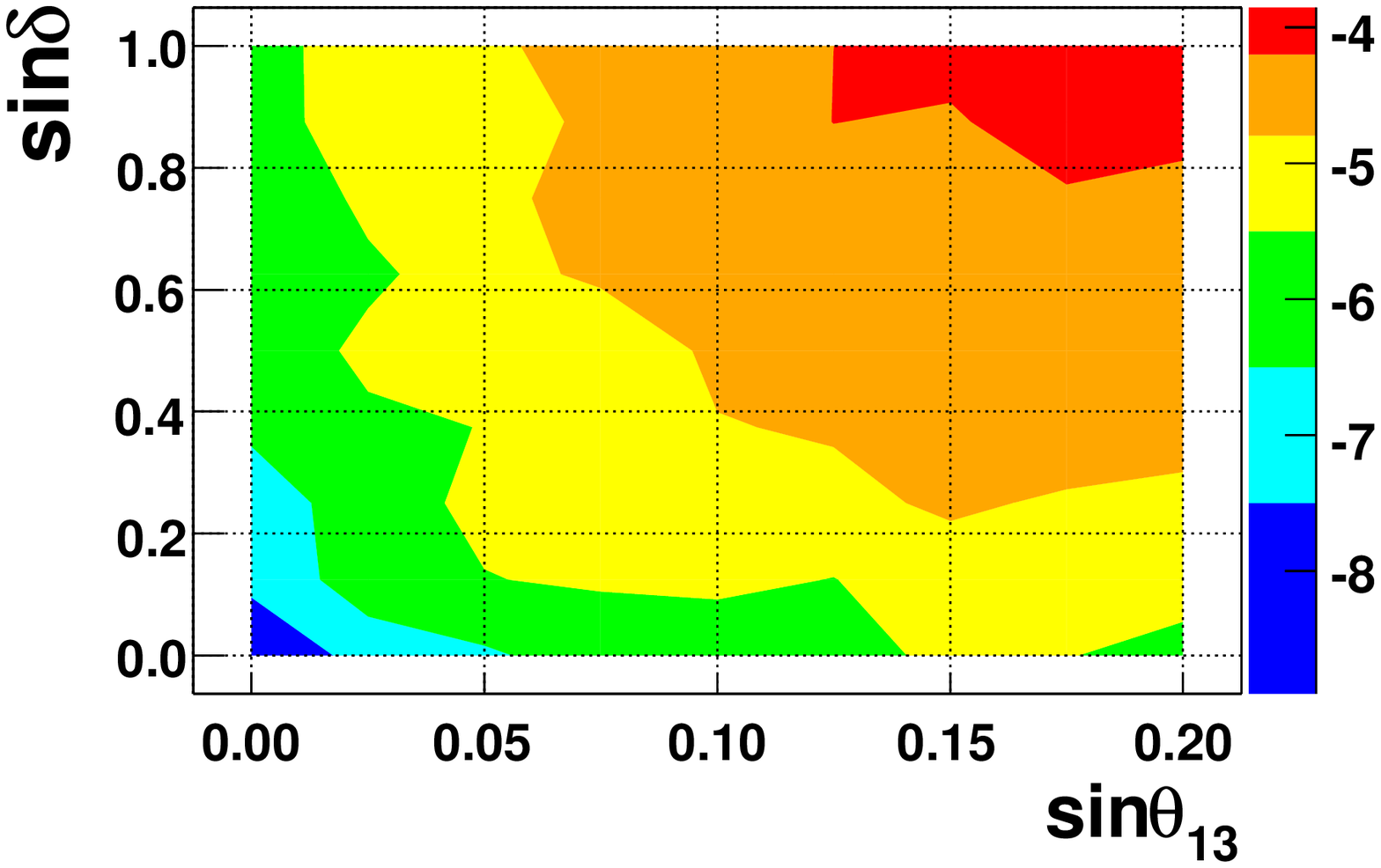}\\    
      \includegraphics[width=0.45\textwidth]{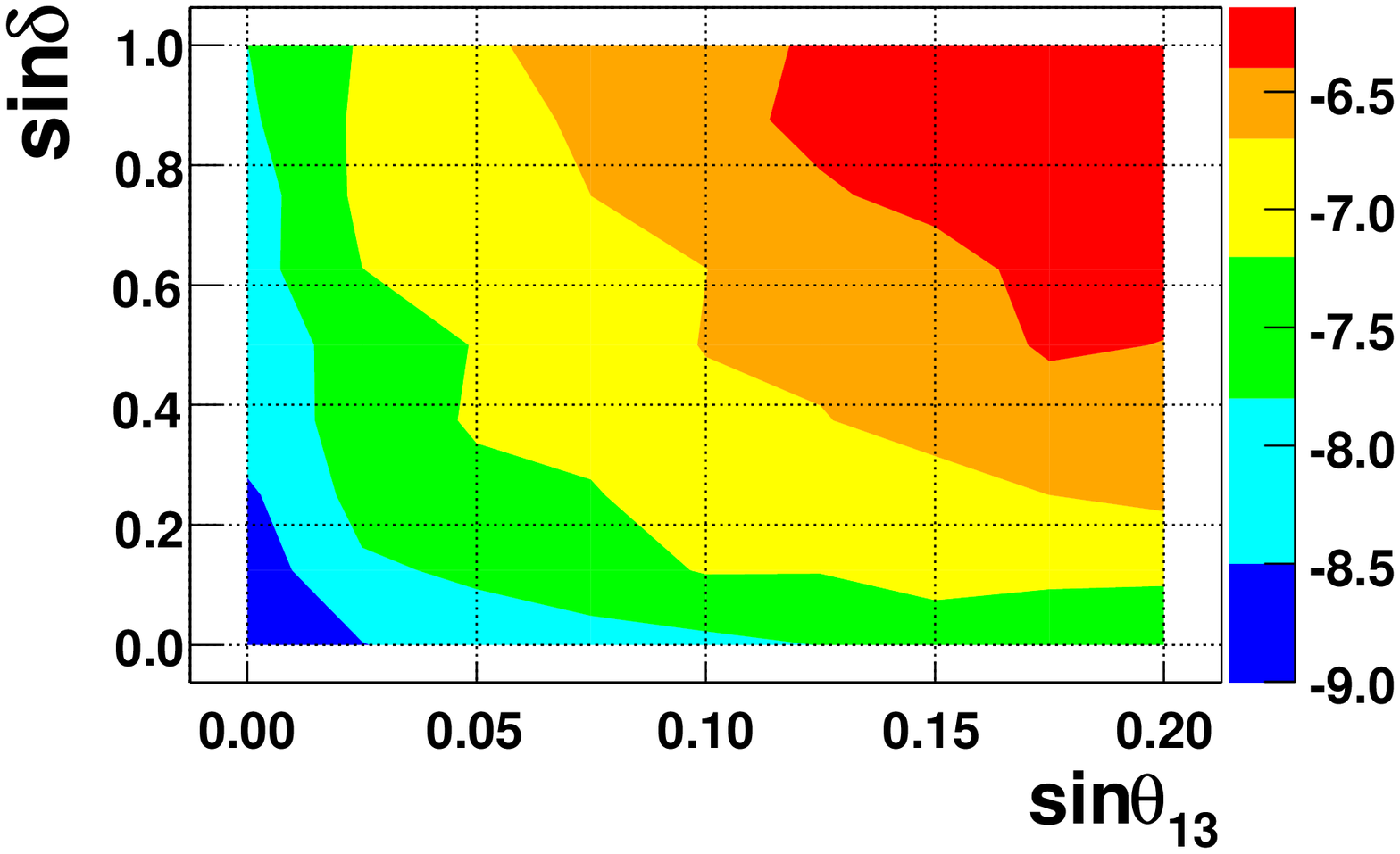} 
    \end{tabular}
  \caption{Contour levels of the maximum value of the asymmetry 
$\epsilon_\Sigma$ obtained as a function of $\sin\theta_{13}$ and
$\sin\delta$. For details see Ref.\cite{romao:2007jr}.} 
    \label{fig:2}
\end{figure}

We proposed\cite{malinsky:2005bi} an alternative inverse seesaw consistent
with a realistic unified \10 model. In this model the mass matrix reads
\begin{equation}
  \label{eq:7}
  M_{\nu} =
  \left(
    \begin{array}{ccc}
      0 & Y v&  F v_{L} \\
      Y^{T}v & 0 & \tilde F v_{R}\\
      F^{T}v_{L} & \tilde F^{T} v_{R}& 0
    \end{array} \right) 
\end{equation}
By inserting $v_{L}=\rho \frac{v_{R} v}{M_{G}}$, coming from the
minimization conditions\cite{malinsky:2005bi}, the $v_{R}$ scale drops
out, leading to
\begin{equation}
  \label{eq:8}
 m_{\rm eff}^{\rm New Inv} \simeq 
    \frac{v^2}{M_{G}} \rho \left[Y( F \tilde F^{-1})^{T}+( F
      \tilde F^{-1}) Y^{T}\right]  
\end{equation}
The neutrino mass is suppressed by $M_{G}$, 
irrespective of how low is the B-L breaking scale $v_{R}$ 
(as low as  few TeV). This corresponds to the following diagram
\begin{center}
\includegraphics[height=27mm]{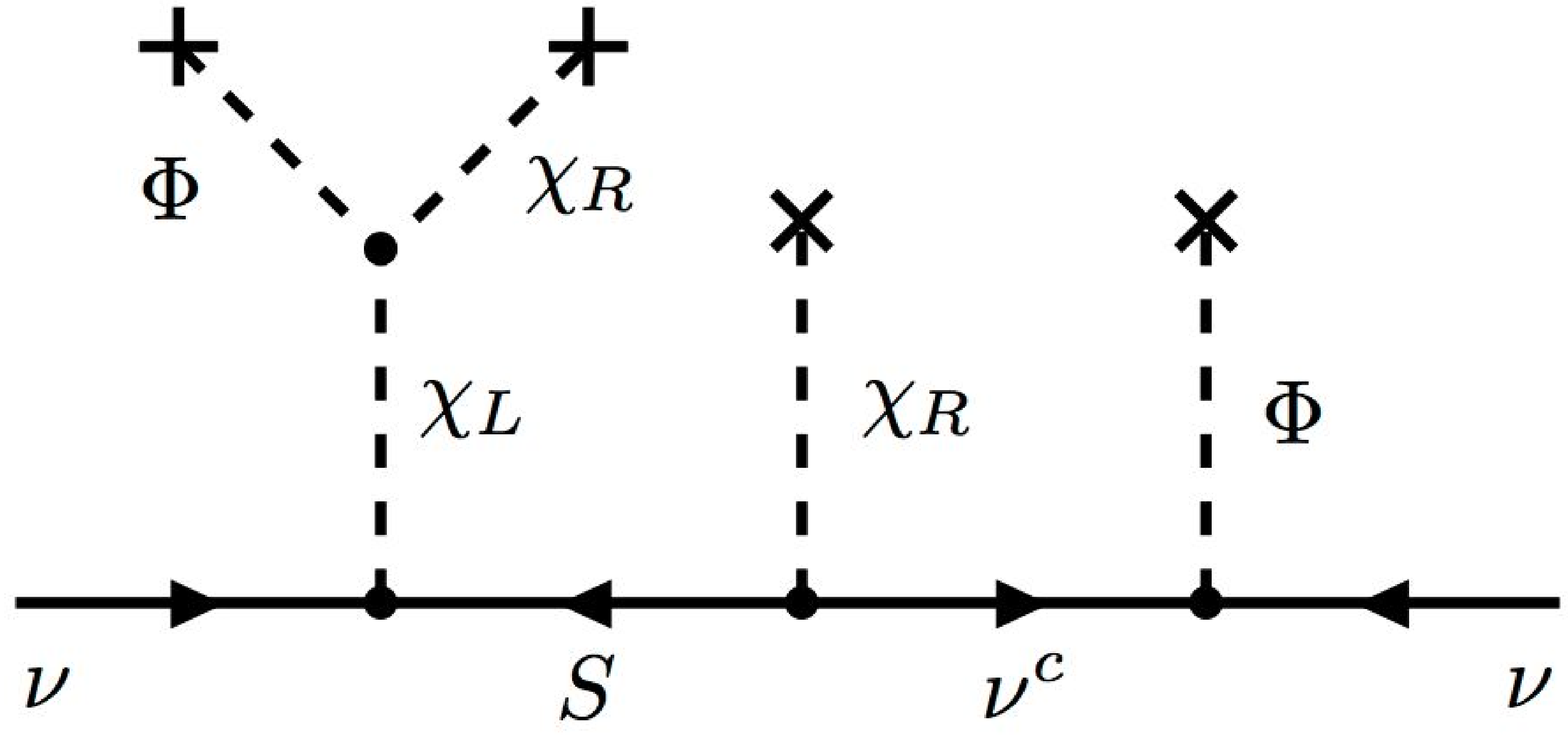} 
\end{center}
This new seesaw is linear in the Dirac Yukawa couplings $Y$. The most
important result is that a low $B-L$ scale can be achieved while
preserving the gauge coupling unification as shown in Fig.~\ref{fig:1}.
Another important result is the calculation of  the CP asymmetry
needed for leptogenesis in a model that is a
variant\cite{hirsch:2006ft} of the previous one. The
results\cite{romao:2007jr} are shown in Fig.~\ref{fig:2}.

\subsection{Models with radiatively generated neutrino mass}

\subsubsection{Zee Model}

  In the Zee model\cite{zee:1980ai}, the scalar sector of the model
  consists of two Higgs doublets with the same hypercharge and a 
  charged $SU(2)$ scalar singlet $h^\pm$, with $L(h^-)=2$.
  \begin{equation}
    \label{eq:9}
    {\cal L}_{Y} = 
              \bar{L}_{i}(\Pi_{a})_{ij}\Phi_{a}e_{Rj} + 
              \epsilon_{\alpha\beta}\bar{L}^{\alpha}_{i}
              f_{ij}\,C(\bar{L}^{T})^{\beta}_{j}h^{-} +
              \hbox{h.c.}
  \end{equation}
Therefore Yukawa interactions conserve $L$, which is explicitly broken
in the scalar potential 
\begin{equation}
  \label{eq:10}
  V\supset\mu\epsilon_{\alpha\beta}\Phi_{1}^{\alpha}
  \Phi_{2}^{\beta}h^{-} + \mbox{h.c.}
\end{equation}
 Charged scalar mixing and Yukawa interactions generate
 neutrino masses at the one loop level through the diagram

 \begin{center}
   \psfrag{n}{$\nu$}
   \psfrag{hp}{$h^+$}
   \psfrag{hm}{$h^-$}
   \psfrag{l}{$l$}
   \psfrag{lb}{$\overline{l}$}
   \includegraphics{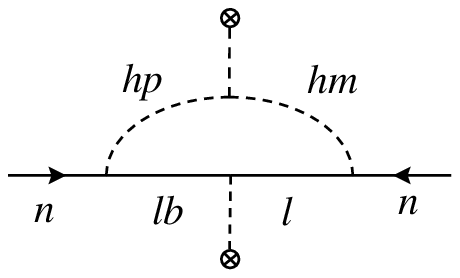}
 \end{center}

%%%-------------------- SLIDE -----------------------
\subsubsection{Babu-Zee Model}
  \vspace{-0.3cm}
 Apart from the Higgs doublet the model\cite{babu:1988ki,zee:1985id},
 contains a single charged and a doubly charged ($h^+$, $k^{++}$) 
 $SU(2)$ gauge singlets scalars. In contrast to the Zee model
 the second Higgs doublet is absent. We have
 \begin{equation}
   \label{eq:11}
   {\cal L} = f_{\alpha\beta} (L^{Ti}_{\alpha L}CL^{j}_{\beta L})
          \epsilon_{ij}h^+
          + h'_{\alpha\beta}(e^T_{\alpha R}Ce_{\beta R})k^{++} + {\rm h.c.}
 \end{equation}
with  $L(h^-)=L(k^{--})=2$. Therefore the Lagrangian conserves $L$,
which is  explicitly broken in the scalar potential
\begin{equation}
  \label{eq:12}
  V\supset\mu k^{++}h^-h^-
\end{equation}
 Yukawa interactions and the breaking of Lepton number, 
generate Majorana neutrino
 masses at the 2-loop level, as indicated in the following diagram: 
\begin{center}
\includegraphics[width=35mm]{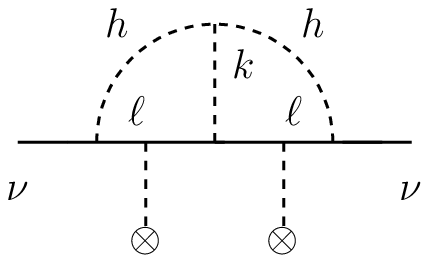}    
\end{center}

%%%-------------------- SLIDE -----------------------
\subsubsection{Broken R-Parity Models}

R--parity is defined in supersymmetric theories by the relation
\begin{equation}
  \label{eq:13}
  R_P=(-1)^{2J +3B +L}
\end{equation}
Although the MSSM is defined to conserve R--parity, there is no
fundamental principle that requires that. If it is not conserved it
can broken in two ways

  \begin{itemize}
  \item  \textbf{BRpV}: \textit{Explicit R-parity Violation}

In this case R--parity is broken at the Lagrangian level. The most
important example is the so-called Bilinear R--parity Violation (BRpV)
model which has the same particle content as the MSSM.

  \item  \textbf{SRpV}: \textit{Spontaneously R-parity Violation}

In this case we have a more complicated Higgs boson structure. The most
salient feature is the existence of the Majoron $J$, the massless
Goldstone boson appearing due to the spontaneous breaking of R--parity.

  \end{itemize}

In the next sections we will review in more detail these two ways
of generating neutrino masses and we will indicate ways of testing
these ideas at accelerators.

%%%-------------------- SLIDE -----------------------

\section{Bilinear R-Parity Violation}

\subsection{The model}

This model\cite{akeroyd:1998iq,diaz:1997xc} has the same particle
content as the MSSM. The superpotential $W$ is 
\begin{eqnarray}
W&=&\varepsilon_{ab}\left[
 h_U^{ij}\widehat Q_i^a\widehat U_j\widehat H_u^b
+h_D^{ij}\widehat Q_i^b\widehat D_j\widehat H_d^a
+h_E^{ij}\widehat L_i^b\widehat R_j\widehat H_d^a \right.\cr
&&\nonumber\\
&&\left. \hskip 1cm 
-\mu\widehat H_d^a\widehat H_u^b
 +\epsilon_i\widehat L_i^a\widehat H_u^b \right]
\end{eqnarray}
where $i,j=1,2,3$ are generation indices, $a,b=1,2$ are $SU(2)$
indices. The set of soft supersymmetry breaking terms are
\begin{equation}
  \label{eq:14}
  V_{soft}=V_{soft}^{\textrm{MSSM}} +\varepsilon_{ab}\
 B_i\epsilon_i\widetilde L^a_i H_u^b
\end{equation}

%%%-------------------- SLIDE -----------------------

\subsection{Tree Level: Atmospheric Mass Scale}

In the basis: $\psi^{0T}= 
(-i\lambda',-i\lambda^3,\widetilde{H}_d^1,\widetilde{H}_u^2,
\nu_{e}, \nu_{\mu}, \nu_{\tau} )$ the mass matrix reads\cite{romao:1999up,hirsch:2000ef}
\begin{equation}
  \label{eq:15}
  {\bold M}_N\hskip -1mm  =\hskip -1mm \left[\hskip -1.5mm  
\begin{array}{cc}  
{\cal M}_{\chi^0}& m^T \cr
\vb{20}
m & 0 \cr
\end{array}\hskip -1mm  
\right]\hskip -1mm  ;
m=\hskip -1mm  \left[  \hskip -1mm  
\begin{array}{cccc}  
-\frac 12g^{\prime }v_1 & \frac 12gv_1 & 0 & \epsilon _1 \cr
-\frac 12g^{\prime }v_2 & \frac 12gv_2 & 0 & \epsilon _2  \cr
-\frac 12g^{\prime }v_3 & \frac 12gv_3 & 0 & \epsilon _3  \cr  
\end{array}  \hskip -1mm  
\right] 
\end{equation}
which gives the effective mass matrix

\begin{eqnarray}
  \label{eq:16}
m_{eff}&\hskip-2mm =&\hskip-2mm - m \, {\cal M}_{\chi^0}^{-1}\, m^T \nonumber\\
&\hskip-2mm=\hskip-2mm& 
\frac{M_1 g^2 + M_2 {g'}^2}{4 \textrm{det}({\cal M}_{\chi^0})} 
\left(\hskip -2mm
\begin{array}{ccc}
\Lambda_e^2 & \Lambda_e \Lambda_\mu
& \Lambda_e \Lambda_\tau \\
\Lambda_e \Lambda_\mu & \Lambda_\mu^2
& \Lambda_\mu \Lambda_\tau \\
\Lambda_e \Lambda_\tau & \Lambda_\mu \Lambda_\tau & \Lambda_\tau^2
\end{array}\hskip -2mm
\right)
  \end{eqnarray}
where $\Lambda_i=\mu v_i + v_d \epsilon_i$. This effective mass matrix
has two massless neutrinos and one massive neutrino 
with mass\cite{romao:1999up}
\begin{equation}
  \label{eq:17}
  m_{\nu} = Tr(m_{eff}) = \displaystyle
\frac{M_1 g^2 + M_2 {g'}^2}{4\, \textrm{det}({\cal M}_{\chi^0})} 
|{\vec \Lambda}|^2.
\end{equation}

\subsection{One Loop: Solar Mass Scale}

\begin{figure}[ht]
  \centering
  \begin{tabular}{cc}
\includegraphics[width=0.45\linewidth]{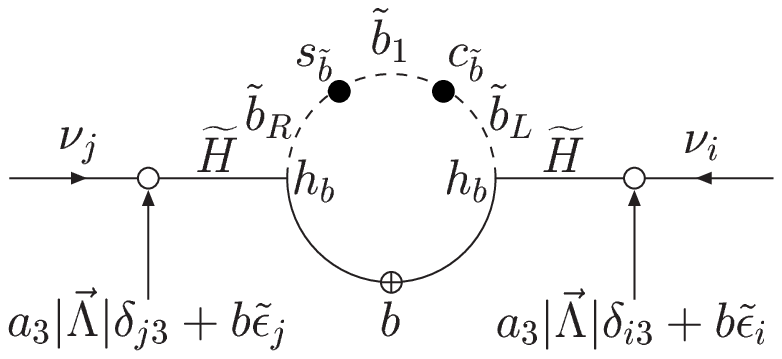}
&\includegraphics[width=0.45\linewidth]{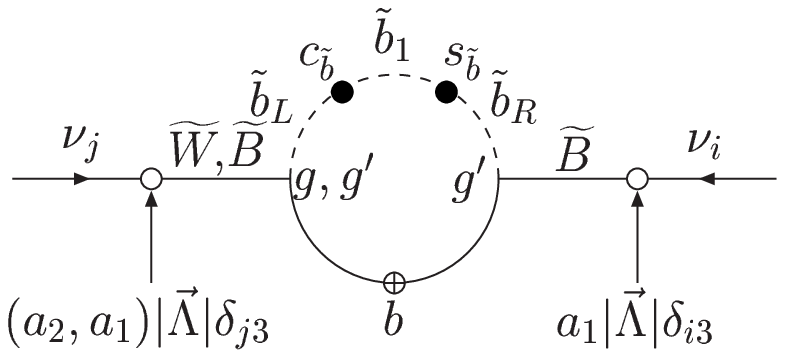}
\end{tabular}
  \caption{Bottom-sbottom loops generating the neutrino masses. In
    this Figure \textit{open circles} correspond to small R-parity
    violating projections, \textit{full circles} correspond to
    R-parity conserving projections and \textit{open circles with a
      cross inside} to  mass insertions which flip chirality.}   
  \label{fig:4}
\end{figure}

\begin{figure}[ht]
  \centering
\includegraphics[width=0.4\textwidth]{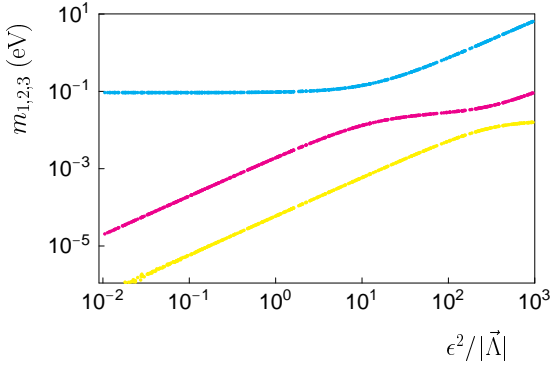}      
  \caption{Neutrino masses in the BRpV for the following values of the
  MSSM parameters:$m_0=\mu=500$ GeV, $\tan \beta= 5$, $B=-A=m_0$. The
  range of BRpV parameters is as follows: $|\vec \Lambda|=0.16$ GeV, $10*
\Lambda_e=\Lambda_{\mu}=\Lambda_{\tau}$ and
$\epsilon_1=\epsilon_2=\epsilon_3$}
  \label{fig:3}
\end{figure}

In this model the solar neutrino mass is generated at one loop
level.
The most important contributions come\cite{diaz:2003as} 
from the bottom-sbottom loops indicated in Fig.~\ref{fig:4}.
In Fig.~\ref{fig:3} we show an example of neutrino masses as function
of the BRpV parameters.

\section{Spontaneous Broken R-Parity}

\subsection{The model }

The most general superpotential that leads to Spontaneous  R-parity
Violation (SRpV) is 
\begin{eqnarray} 
\label{eq:W}
{\cal W} &\hskip-4mm=\hskip-4mm& \varepsilon_{ab}\Big(
h_U^{ij}\widehat Q_i^a\widehat U_j\widehat H_u^b 
+h_D^{ij}\widehat Q_i^b\widehat D_j\widehat H_d^a 
+h_E^{ij}\widehat L_i^b\widehat E_j\widehat H_d^a \nonumber \\
& &
+h_{\nu}^{ij}\widehat L_i^a\widehat \nu^c_j\widehat H_u^b 
\!- {\hat \mu}\widehat H_d^a\widehat H_u^b 
\!- (h_0 \widehat H_d^a\widehat H_u^b +\delta^2)\widehat\Phi \Big) 
\nonumber \\
& &+  h^{ij} \widehat\Phi \widehat\nu^c_i\widehat S_j +
M_{R}^{ij}\widehat \nu^c_i\widehat S_j 
+ \frac{1}{2}M_{\Phi} \widehat\Phi^2 +\frac{\lambda}{3!}
\widehat\Phi^3 
\end{eqnarray}
where the  singlet superfields $({\nu^c}_i,S_i,\Phi)$
  carry a conserved lepton number assigned as $(-1, 1,0)$. Therefore 
Lepton number and R-parity are conserved at Lagrangian Level.

%%%-------------------- SLIDE -----------------------
\subsection{SRpV: Symmetry Breaking}

    The spontaneous breaking of R--parity is driven by nonzero vevs for
    the scalar neutrinos. The scale characterizing R--parity breaking is
    set by the isosinglet vevs 
\begin{equation}
  \label{eq:18}
  \vev{\tilde{\nu^c}} = \frac{v_R}{\sqrt{2}},\quad
  \vev{\tilde{S}}=\frac{v_S}{\sqrt{2}},\quad 
  \vev{\Phi} = \frac{v_{\Phi}}{\sqrt{2}}
\end{equation}
 We also have  very small left-handed sneutrino vacuum expectation values
 \begin{equation}
   \label{eq:19}
          \vev{\tilde{\nu}_{Li}} = \frac{v_{Li}}{\sqrt{2}}
 \end{equation}
The electroweak breaking is driven by 
\begin{equation}
  \label{eq:20}
\vev{{H_u}} = \frac{v_{u}}{\sqrt{2}},\quad \vev{{H_d}}=
    \frac{v_{d}}{\sqrt{2}}
  \end{equation}
with $v^2 = v_u^2 + v_d^2 + \sum_i v_{L i}^2$ and $m_W^2 = \frac{g^2 v^2}{4}$.
The spontaneous breaking of R--parity also entails the spontaneous
violation of total lepton number. This implies that the \textit{Majoron}
\begin{equation}
  \label{eq:21}
    J=\mathrm{Im} \left[\displaystyle
      \frac{v_L^2}{Vv^2} (v_u H_u - v_d H_d) +
      \sum_i \frac{v_{Li}}{V} \tilde{\nu_{i}} 
      +\frac{v_S}{V} S
      -\frac{v_R}{V} \tilde{\nu^c}\right]  
\end{equation}
remains massless, as it is the Nambu-Goldstone boson associated to the
breaking of lepton number.

%%%-------------------- SLIDE -----------------------
\subsection{SRpV: The effective neutrino mass matrix}

The effective neutrino mass 
matrix can be cast into a very simple form 
\begin{equation}
  \label{eq:22}
(\mathbf{m_{\nu\nu}^{\rm eff}})_{ij} = a \Lambda_i \Lambda_j + 
    b (\epsilon_i \Lambda_j + \epsilon_j \Lambda_i) +
    c \epsilon_i \epsilon_j.  
\end{equation}
This equation resembles very closely the result for the
{BRpV} model once the
dominant 1-loop corrections are taken into account.
The effective bilinear R--parity violating
parameters are
\begin{equation}
  \label{eq:23}
  \epsilon_{i} = h_{\nu}^{i}\, \frac{v_R}{\sqrt{2}}
\quad \hbox{and} \quad
\Lambda_i = \epsilon_i v_d + \mu v_{L_i}
\end{equation}

\section{Tests of the Bilinear R-Parity Violation Model}

\subsection{Testing BRpV via SUSY Decays}

\begin{itemize}
\item 
{LSP Decays:}
{\bf (mSUGRA)}

\vspace{1mm}
The fact that, in these models, the LSP decays through R--parity 
violating processes allows it to
be either neutral or charged. 

\begin{itemize}
\item 
In most cases the LSP is the lightest neutralino, like in the
MSSM\cite{porod:2000hv}. 

\item
For some regions of the parameter space the LSP can also be the scalar
tau\cite{hirsch:2002ys}.

\item
In both cases we have shown that despite the smallness of $m_{\nu}$
the LSP decays inside the detector.

\end{itemize}
\item 
{LSP Decays:} {\bf
(non mSUGRA)}

\vspace{1mm}
If we depart from mSUGRA then the LSP can be almost any
particle\cite{hirsch:2003fe}.
This gives complementary information. 

\end{itemize}

%%%-------------------- SLIDE -----------------------

\subsection{Neutralino decays: Probing the Atmospheric Angle}

\begin{figure}[ht]
  \centering
\includegraphics[width=0.37\textwidth]{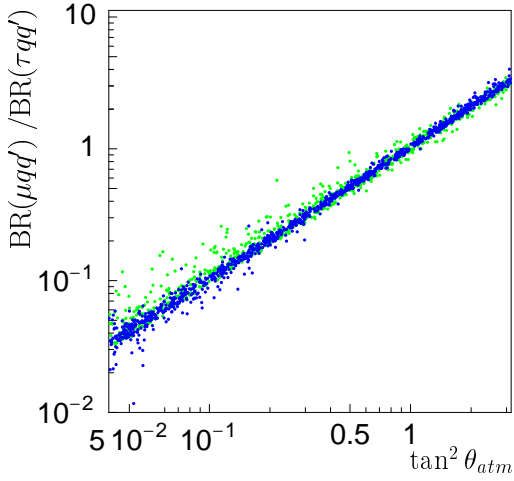}  
  \caption{Relation of branching of branching ratios of neutralino
    decays and the atmospheric angle (Ref.~\cite{porod:2000hv}). }
  \label{fig:5}
\end{figure}

\begin{figure}[ht]
  \centering
\includegraphics[width=0.37\textwidth]{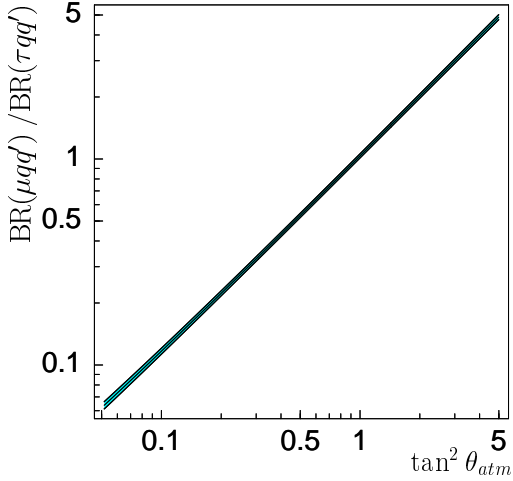}  
  \caption{Same as in Fig.~\ref{fig:5}, but
    assuming 
    that SUSY has been
    discovered with $M_2=120\, \hbox{GeV}, \mu=500\, \hbox{GeV},
    \tan\beta=5$ $ m_{0}=500\, \hbox{GeV}, A=-500\, \hbox{GeV}$
    (Ref.~\cite{porod:2000hv}).} 
  \label{fig:6}
\end{figure}

When the LSP is the neutralino the ratios of branching ratios can be
correlated with the neutrino parameters. For instance, in
Fig. \ref{fig:5} we show\cite{porod:2000hv} the correlation with 
the atmospheric angle. 
The spread in Fig.~\ref{fig:5} will disappear if the SUSY parameters
were known, as it is indicated in Fig.~\ref{fig:6}.

%%%-------------------- SLIDE -----------------------

\subsection{Stau Decays  and the Solar Angle}

\begin{figure}[ht]
  \centering
\includegraphics[width=0.4\textwidth,height=50mm]{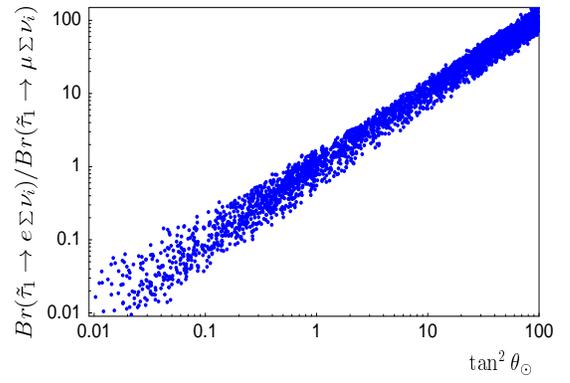}  
  \caption{Correlations of the stau decays with the solar angle
    (Ref.~\cite{hirsch:2002ys}).}
  \label{fig:7}
\end{figure}

In the region of parameters space where the LSP is the stau we can use
its decays\cite{hirsch:2002ys}, 
to make correlations to the neutrino parameters as is indicated in
Fig.~\ref{fig:7}.  

%%%-------------------- SLIDE -----------------------

\subsection{Other LSP Decays (depart from {\bf mSUGRA})}

\begin{figure}[ht]
  \centering
\includegraphics[width=0.4\textwidth]{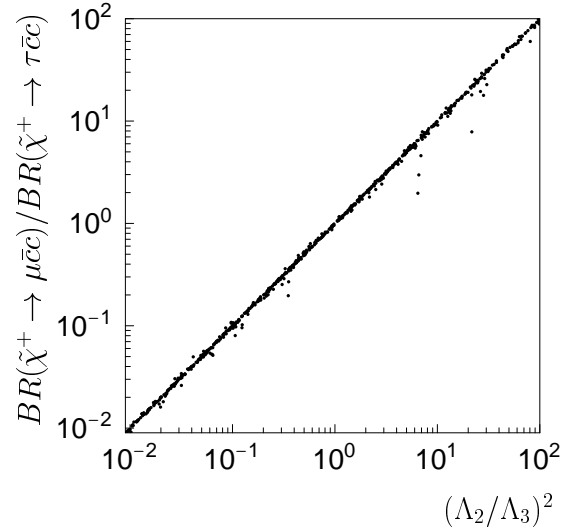}  
  \caption{Correlations for the atmospheric parameters when the LSP is
    the chargino (Ref.~\cite{hirsch:2003fe}).}
  \label{fig:8}
\end{figure}

\begin{figure}[ht]
  \centering
\includegraphics[width=0.4\textwidth]{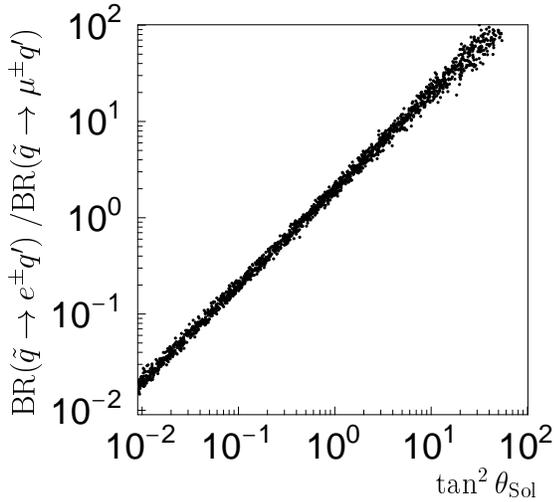}  
  \caption{Correlations for the solar angle when the LSP is the squark
    (Ref.~\cite{hirsch:2003fe}).}
  \label{fig:9}
\end{figure}

If we depart from mSUGRA then the LSP can be almost any particle. For
instance, it was shown in Ref.\cite{hirsch:2003fe} 
that you can correlate the decays of charginos and squarks with the
neutrino parameters. In Fig.~\ref{fig:8} and Fig.~\ref{fig:9} we show,
respectively the case of the LSP being the chargino or the squark.

\section{Tests of the Spontaneous R-Parity Violation Model}

%%%-------------------- SLIDE -----------------------
\subsection{Higgs Boson Production}

    Supersymmetric Higgs bosons can be produced at the $e^+
     e^-$ collider via the so--called  Bjorken process, coming from
     the coupling
\begin{equation}
  \label{eq:24}
  {\cal L}_{HZZ}
  =\displaystyle
  \sum_{i=1}^8 (\sqrt 2 G_F)^{1/2} M_Z^2 Z_{\mu}Z^{\mu} \eta_{i} H_i
\end{equation}
where the $\eta_i$ are a combination of the doublet scalars,
\begin{equation}
  \label{eq:25}
  \eta_i= \frac{v_d}{v} \
  R^S_{i 1}
  + \frac{v_u}{v} \
  R^S_{i 2} 
  + \sum_{j=1}^3 \frac{v_{Lj}}{v} \
  R^S_{i j+2}
\end{equation}
and the $R^S_{i j}$ are rotation matrices that diagonalize the CP even
neutral Higgs bosons.
In comparison with the SM, the coupling of the lightest CP--even Higgs
boson to the $Z$ is reduced by
\begin{equation}
  \label{eq:26}
  \eta \equiv \eta_1 \leq 1
\end{equation}

%%%-------------------- SLIDE -----------------------
\subsection{Higgs Boson Decay}

We are interested here in the ratio
\begin{equation}
  \label{eq:27}
    \label{eq:ratio}
    R_{Jb}=\displaystyle
    \frac{\Gamma(h \to JJ)}{\Gamma(h \to b \bar{b})}  
\end{equation}
of the invisible decay to the SM decay into b-jets. These decay widths are
\begin{equation}
  \label{eq:28}
    \Gamma(h\to  JJ)=\displaystyle
    \frac{g_{hJJ}^2}{32\pi m_h}   
\end{equation}
\vspace{3mm}
and
\begin{equation}
  \label{eq:29}
\hskip -1mm
    \Gamma(h\to  b \bar{b})=
    \frac{3 G_F \sqrt{2}}{8\pi}\!\!
    \left(R^S_{11}
        \right)^2\! m_h  m_b^2 \left[\! 1\!-\!4
  \left(\frac{m_b}{m_h}\right)^2 \! \right]^{3/2}
  \end{equation}

%%%-------------------- SLIDE -----------------------

\subsection{Numerical results} 

\begin{figure}[ht]
  \centering
  \includegraphics[clip,width=0.45\textwidth]{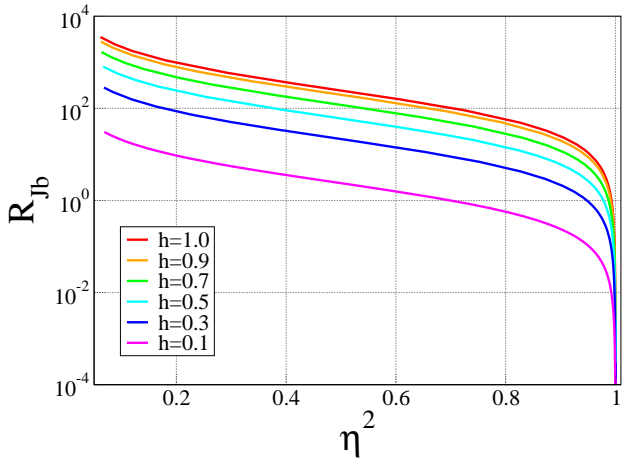}  
  \caption{$R_{Jb}$ vs $\eta^2$ for fixed $h$ from
    Ref.\cite{hirsch:2004rw}. See text, Eq.~(\ref{eq:25}), 
    Eq.~(\ref{eq:27}) and Eq.~(\ref{eq:W}) for an explanation of these
    quantities. }
  \label{fig:10}
\end{figure}

We have performed a careful analysis of the SBpV model in order to
search for the possibility of having at the same time a large
branching ratio of the Higgs boson into the invisible channel of the
Majoron and at the same time the production cross section to be not
too much reduced with respect to the SM case. The
conclusion\cite{hirsch:2004rw,hirsch:2005wd} is that
this is indeed possible as is shown, as an example in
Fig.~\ref{fig:10}. All the points in these curves respect the neutrino
data\cite{maltoni:2004ei}.

\section{Conclusions}
%%%-------------------- SLIDE -----------------------

  We have briefly reviewed the models for neutrino mass. In the class
  of seesaw models we discussed in some detail the implications of a
  new seesaw model that can achieve successful leptogenesis and can
  have low energy implications.  Among the radiative models we have
  focused on those with broken R-parity. We have considered both the
  models with explicit R-parity violation (BRpV) and also models with
  spontaneous violation of R-parity (SRpV). Both possibilities have
  implications at the new accelerators and therefore can be tested in
  the upcoming machines.

% use section* for acknowledgement
\section*{Acknowledgment}

Work supported by European Commission Contracts
MRTN-CT-2004-503369 and ILIAS/N6 WP1 RII3-CT-2004-506222.

%\newpage

%\bibliographystyle{h-physrev4} 
%\bibliography{romao-ref}

\begin{thebibliography}{10}

\bibitem{fukuda:1998mi}
Super-Kamiokande, Y.~Fukuda {\em et~al.},
\newblock Phys. Rev. Lett. {\bf 81}, 1562 (1998), [hep-ex/9807003].
%%CITATION = HEP-EX 9807003;%%



\bibitem{ahmad:2002jz}
SNO, Q.~R. Ahmad {\em et~al.},
\newblock Phys. Rev. Lett. {\bf 89}, 011301 (2002), [nucl-ex/0204008].
%%CITATION = NUCL-EX 0204008;%%



\bibitem{eguchi:2002dm}
KamLAND, K.~Eguchi {\em et~al.},
\newblock Phys. Rev. Lett. {\bf 90}, 021802 (2003), [hep-ex/0212021].
%%CITATION = HEP-EX 0212021;%%

\bibitem{maltoni:2004ei}
M.~Maltoni, T.~Schwetz, M.~A. Tortola and J.~W.~F. Valle,
\newblock New J. Phys. {\bf 6}, 122 (2004), [hep-ph/0405172].
%%CITATION = HEP-PH 0405172;%%

\bibitem{minkowski:1977sc}
P.~Minkowski,
\newblock Phys. Lett. {\bf B67}, 421 (1977).
%%CITATION = PHLTA,B67,421;%%

\bibitem{gell-mann:1980vs}
M.~Gell-Mann, P.~Ramond and R.~Slansky,
\newblock (1979),
\newblock Print-80-0576 (CERN).

\bibitem{yanagida:1979}
T.~Yanagida,
\newblock (1979),
\newblock ed. Sawada and Sugamoto (KEK, 1979).

\bibitem{mohapatra:1981yp}
R.~N. Mohapatra and G.~Senjanovic,
\newblock Phys. Rev. {\bf D23}, 165 (1981).
%%CITATION = PHRVA,D23,165;%%

\bibitem{chikashige:1981ui}
Y.~Chikashige, R.~N. Mohapatra and R.~D. Peccei,
\newblock Phys. Lett. {\bf B98}, 265 (1981).
%%CITATION = PHLTA,B98,265;%%

\bibitem{schechter:1980gr}
J.~Schechter and J.~W.~F. Valle,
\newblock Phys. Rev. {\bf D22}, 2227 (1980).
%%CITATION = PHRVA,D22,2227;%%

\bibitem{lazarides:1980nt}
G.~Lazarides, Q.~Shafi and C.~Wetterich,
\newblock Nucl. Phys. {\bf B181}, 287 (1981).
%%CITATION = NUPHA,B181,287;%%

\bibitem{malinsky:2005bi}
M.~Malinsky, J.~C. Romao and J.~W.~F. Valle,
\newblock Phys. Rev. Lett. {\bf 95}, 161801 (2005), [hep-ph/0506296].
%%CITATION = HEP-PH 0506296;%%

\bibitem{mohapatra:1986bd}
R.~N. Mohapatra and J.~W.~F. Valle,
\newblock Phys. Rev. {\bf D34}, 1642 (1986).
%%CITATION = PHRVA,D34,1642;%%

\bibitem{zee:1980ai}
A.~Zee,
\newblock Phys. Lett. {\bf B93}, 389 (1980).
%%CITATION = PHLTA,B93,389;%%

\bibitem{babu:1988ki}
K.~S. Babu,
\newblock Phys. Lett. {\bf B203}, 132 (1988).
%%CITATION = PHLTA,B203,132;%%

\bibitem{aulakh:1982yn}
C.~S. Aulakh and R.~N. Mohapatra,
\newblock Phys. Lett. {\bf B119}, 136 (1982).
%%CITATION = PHLTA,B119,136;%%

\bibitem{diaz:1998xc}
M.~A. Diaz, J.~C. Romao and J.~W.~F. Valle,
\newblock Nucl. Phys. {\bf B524}, 23 (1998), [hep-ph/9706315].
%%CITATION = HEP-PH 9706315;%%

\bibitem{diaz:2003as}
M.~A. Diaz, M.~Hirsch, W.~Porod, J.~C. Romao and J.~W.~F. Valle,
\newblock Phys. Rev. {\bf D68}, 013009 (2003), [hep-ph/0302021].
%%CITATION = HEP-PH 0302021;%%

\bibitem{romao:1999up}
J.~C. Romao, M.~A. Diaz, M.~Hirsch, W.~Porod and J.~W.~F. Valle,
\newblock Phys. Rev. {\bf D61}, 071703 (2000), [hep-ph/9907499].
%%CITATION = HEP-PH 9907499;%%

\bibitem{hirsch:2000ef}
M.~Hirsch, M.~A. Diaz, W.~Porod, J.~C. Romao and J.~W.~F. Valle,
\newblock Phys. Rev. {\bf D62}, 113008 (2000), [hep-ph/0004115].
%%CITATION = HEP-PH 0004115;%%

\bibitem{hirsch:2004he}
M.~Hirsch and J.~W.~F. Valle,
\newblock New J. Phys. {\bf 6}, 76 (2004), [hep-ph/0405015].
%%CITATION = HEP-PH 0405015;%%

\bibitem{hirsch:2002ys}
M.~Hirsch, W.~Porod, J.~C. Romao and J.~W.~F. Valle,
\newblock Phys. Rev. {\bf D66}, 095006 (2002), [hep-ph/0207334].
%%CITATION = HEP-PH 0207334;%%

\bibitem{porod:2000hv}
W.~Porod, M.~Hirsch, J.~Romao and J.~W.~F. Valle,
\newblock Phys. Rev. {\bf D63}, 115004 (2001), [hep-ph/0011248].
%%CITATION = HEP-PH 0011248;%%

\bibitem{restrepo:2001me}
D.~Restrepo, W.~Porod and J.~W.~F. Valle,
\newblock Phys. Rev. {\bf D64}, 055011 (2001), [hep-ph/0104040].
%%CITATION = HEP-PH 0104040;%%

\bibitem{hirsch:2003fe}
M.~Hirsch and W.~Porod,
\newblock Phys. Rev. {\bf D68}, 115007 (2003), [hep-ph/0307364].
%%CITATION = HEP-PH 0307364;%%

\bibitem{masiero:1990uj}
A.~Masiero and J.~W.~F. Valle,
\newblock Phys. Lett. {\bf B251}, 273 (1990).
%%CITATION = PHLTA,B251,273;%%

\bibitem{romao:1992vu}
J.~C. Romao, C.~A. Santos and J.~W.~F. Valle,
\newblock Phys. Lett. {\bf B288}, 311 (1992).
%%CITATION = PHLTA,B288,311;%%

\bibitem{shiraishi:1993di}
M.~Shiraishi, I.~Umemura and K.~Yamamoto,
\newblock Phys. Lett. {\bf B313}, 89 (1993).
%%CITATION = PHLTA,B313,89;%%

\bibitem{romao:1992ex}
J.~C. Romao and J.~W.~F. Valle,
\newblock Nucl. Phys. {\bf B381}, 87 (1992).
%%CITATION = NUPHA,B381,87;%%

\bibitem{romao:1992zx}
J.~C. Romao, F.~de~Campos and J.~W.~F. Valle,
\newblock Phys. Lett. {\bf B292}, 329 (1992), [hep-ph/9207269].
%%CITATION = HEP-PH 9207269;%%

\bibitem{decampos:1994rw}
F.~De~Campos, J.~W.~F. Valle, A.~Lopez-Fernandez and J.~C. Romao,
\newblock hep-ph/9405382.
%%CITATION = HEP-PH 9405382;%%

\bibitem{decampos:1996bg}
F.~de~Campos, O.~J.~P. Eboli, J.~Rosiek and J.~W.~F. Valle,
\newblock Phys. Rev. {\bf D55}, 1316 (1997), [hep-ph/9601269].
%%CITATION = HEP-PH 9601269;%%

\bibitem{hirsch:2004rw}
M.~Hirsch, J.~C. Romao, J.~W.~F. Valle and A.~Villanova~del Moral,
\newblock Phys. Rev. {\bf D70}, 073012 (2004), [hep-ph/0407269].
%%CITATION = HEP-PH 0407269;%%

\bibitem{hirsch:2005wd}
M.~Hirsch, J.~C. Romao, J.~W.~F. Valle and A.~Villanova~del Moral,
\newblock Phys. Rev. {\bf D73}, 055007 (2006), [hep-ph/0512257].
%%CITATION = HEP-PH 0512257;%%

\bibitem{weinberg:1980bf}
S.~Weinberg,
\newblock Phys. Rev. {\bf D22}, 1694 (1980).
%%CITATION = PHRVA,D22,1694;%%

\bibitem{schechter:1981cv}
J.~Schechter and J.~W.~F. Valle,
\newblock Phys. Rev. {\bf D25}, 774 (1982).
%%CITATION = PHRVA,D25,774;%%

\bibitem{hirsch:2006ft}
M.~Hirsch, J.~W.~F. Valle, M.~Malinsky, J.~C. Romao and U.~Sarkar,
\newblock Phys. Rev. {\bf D75}, 011701 (2007), [hep-ph/0608006].
%%CITATION = HEP-PH/0608006;%%

\bibitem{romao:2007jr}
J.~C. Romao, M.~A. Tortola, M.~Hirsch and J.~W.~F. Valle,
\newblock arXiv:0707.2942 [hep-ph].
%%CITATION = ARXIV:0707.2942;%%

\bibitem{zee:1985id}
A.~Zee,
\newblock Nucl. Phys. {\bf B264}, 99 (1986).
%%CITATION = NUPHA,B264,99;%%

\bibitem{akeroyd:1998iq}
A.~G. Akeroyd, M.~A. Diaz, J.~Ferrandis, M.~A. Garcia-Jareno and J.~W.~F.
  Valle,
\newblock Nucl. Phys. {\bf B529}, 3 (1998), [hep-ph/9707395].
%%CITATION = HEP-PH 9707395;%%

\bibitem{diaz:1997xc}
M.~A. Diaz, J.~C. Romao and J.~W.~F. Valle,
\newblock Nucl. Phys. {\bf B524}, 23 (1998), [hep-ph/9706315].
%%CITATION = HEP-PH 9706315;%%

\end{thebibliography}

\end{document}